\documentclass[12pt]{article}
\usepackage{amssymb,amsmath,epsfig}
\begin{document} \parindent=0pt
\parskip=6pt \rm \vspace*{0.5cm}
\begin{center}
 {\bf \Large Phase
diagram of a class of spin-triplet\\
ferromagnetic superconductors}

{\bf  D. V. Shopova$^{\ast}$ and D. I. Uzunov}

{\em  CPCM Laboratory, G. Nadjakov
Institute of Solid State Physics,\\
Bulgarian Academy of Sciences, BG-1784
Sofia, Bulgaria.} \\ \end{center}

$^{\ast}$ Corresponding author: sho@issp.bas.bg

\vspace{0.5cm}

order parameter profile.


\vspace{0.3cm}
\normalsize

\begin{abstract}
 We investigate thermodynamic phases, including the
 phase of coexistence of superconductivity and ferromagnetism, the possible
 phase transitions of first and second order, and the shape of the
phase diagram  in mean-field approximation for a phenomenological
 model of spin-triplet ferromagnetic superconductors. The results are
 discussed in view of application to metallic ferromagnets as UGe$_2$,
 ZrZn$_2$,  URhGe, and Fe.
\end{abstract}

Recently, the coexistence of ferromagnetism and superconductivity was
 discovered in the metallic compounds
 UGe$_2$~\cite{Saxena:2000,Huxley:2001,Tateiwa:2001},
 ZrZn$_2$~\cite{Pfleiderer:2001}, URhGe~\cite{Aoki:2001} and also
 in Fe~\cite{Shimizu:2001} in experiments at low temperatures and
 high pressure.
 In contrast to other superconducting materials, in these
metals the phase transition temperature to the ferromagnetic state is
 higher than the
 phase transition temperature to the superconducting state. Moreover,
it seems that the superconductivity in the metallic compounds mentioned
above always coexists with the ferromagnetic order and is
 enhanced by the latter. In these systems the superconductivity seems to arise
 from the same electrons that create the band magnetism, and is
 most naturally understood as a triplet rather than spin-singlet pairing
 phenomenon~\cite{Volovik:1985, Sigrist:1991,Mineev:1999}.
 The same unconventional superconductivity has been suggested~\cite{Saxena:2000}
 as a possible outcome of the interpretation of  experiments in
  Fe ~\cite{Shimizu:2001}. Note,
 that both vortex and Meissner superconductivity
 phases~\cite{Shimizu:2001} are found in the
  high-pressure crystal modification of Fe which has a
 hexagonal close-packed lattice. In this hexagonal lattice the strong
 ferromagnetism of the usual bcc iron crystal probably
 disappears~\cite{Saxena:2001}.

Recently, the  phenomenological theory that explains coexistence of
ferromagnetism and unconventional spin-triplet superconductivity of
Landau-Ginzburg type was developed ~\cite{Machida:2001, Walker:2002}.
The possible low-order coupling between the superconducting and
ferromagnetic order parameters is derived on the basis of general
symmetry  group arguments and several important features of the
superconducting vortex state in the ferromagnetic phase of
unconventional ferromagnetic superconductors are
established~\cite{Machida:2001, Walker:2002}.

In this letter we shall use the approach presented
 in Refs.~\cite{Machida:2001, Walker:2002}  to investigate
 the conditions
 for the occurrence of the Meissner phase and to demonstrate that the presence
of ferromagnetic order enhances the $p$-wave superconductivity.
 For this aim we shall establish
 the phase diagram corresponding to  model ferromagnetic
superconductors in a zero external magnetic field. We shall also show
that the phase transition to the superconducting state in ferromagnetic
superconductors can be either of first or  second order depending on
the particular substance. We confirm the predictions made in
Refs.~\cite{Machida:2001,Walker:2002} about the symmetry of the ordered
phases.

Our investigation is based on the mean-field
 approximation~\cite{Uzunov:1993} as well as on familiar results
about the possible phases in nonmagnetic superconductors with triplet
($p$-wave) Cooper pairs~\cite{Volovik:1985, Blagoeva:1990,
Uzunov:1990}. We shall neglect all anisotropies,  usually given by
 the respective additional Landau invariants and gradient
 terms~\cite{Sigrist:1991, Mineev:1999} in the Ginzburg-Landau free energy of
 unconventional superconductors.
 The reasons is that the inclusion of  crystal anisotropy is
 related with lengthy formulae and a multivariant analysis which will
 obscure our main aims and results. Let us emphasize that the present results
 should be valid in the same or modified form
 when the crystal anisotropy is properly taken into account. We have to mention also that there
  is a formal similarity between the phase diagram obtained in our
investigation and the phase diagram of certain improper
ferroelectrics~\cite{Gufan:1980,Gufan:1986}.

Following Refs.~\cite{Machida:2001, Walker:2002}
 we consider the Ginzburg-
 Landau free energy $F=\int d^3 x f(\psi, \vec{{\cal{M}}})$, where
\begin{equation}
\label{eq1}
f = \frac{\hbar^2}{4m} (D^{\ast}_j\psi)(D_j\psi)
+ a_s|\psi|^2 + \frac{b}{2}|\psi|^4
+ a_f\vec{{\cal{M}}}^2 +
\frac{\beta}{2}{\cal{M}}^4 +
i\gamma_0 \vec{{\cal{M}}}.(\psi\times \psi^*)\;.
\end{equation}
In Eq.~(\ref{eq1}), $D_j =(\nabla - 2ieA_j/\hbar c)$, where $A_j$ ($j =
1,2,3$)
 are the components of the vector potential $\vec{A}$ related
with the magnetic induction $\vec{B} = \nabla \times \vec{A}$,
the complex vector
 $\psi = \{\psi_j\} \equiv \left(\psi_1,\psi_2,\psi_3\right)$ is the superconducting
 order parameter, corresponding to the spin-triplet Cooper pairing
and $\vec{{\cal{M}}}= \{{\cal{M}}_j\}$ is the magnetization. The
coupling constant
 $\gamma_0 = 4\pi J>0$ is given by the ferromagnetic exchange parameter
 $(J>0)$. Coefficients $a_s = \alpha_s(T-T_s)$ and $a_f = \alpha_f(T-T_f)$
 are expressed by the positive
 parameters $\alpha_s$ and $\alpha_f$ as well as by the
 superconducting $(T_s)$ and ferromagnetic $(T_f)$ critical
temperatures
 in the decoupled case, when
 ${\cal{M}}\psi$-interaction is ignored;
 $b > 0$ and $\beta > 0$ as usual. Note, that all seven material parameters ($\alpha_s$,
 $\alpha_f$, $T_s$, $T_s$, $b$, $\beta$, $J$) depend on material properties,
 the temperature $T$ and
 additional intensive thermodynamic parameters as pressure $P$.

We assume that the magnetization ${\cal{M}}$ is uniform, which is a
reliable assumption outside a quite close vicinity of the
 magnetic phase transition whereas we keep the spatial ($\vec{x}-$)
 dependence of $\psi$. The reason is that the relevant dependence
 of $\psi$ on $\vec{x}$ is generated by the diamagnetic effects
 arising from the presence of ${\cal{M}}$ and the external
 magnetic field $\vec{H}$~\cite{Machida:2001,Walker:2002} rather than
 from fluctuations of $ \psi $ (this effect is extremely small and
 can be safely ignored). Note, that the first term in~(\ref{eq1})
still
 persists for $\vec{H} = 0$ because of the diamagnetic effect created by
 of magnetization $\vec{{\cal{M}}} = \vec{B}/4\pi > 0$. As we shall
 investigate the conditions for the occurrence of the
 Meissner phase where $\psi$ is uniform, the
 spatial dependence of $\psi$ and, hence, the first term in  r.h.s.
of~(\ref{eq1}) will be neglected (see also a brief discussion at the end of
this paper).

In case of a strong easy axis type of magnetic anisotropy, as is
 in UGe$_2$~\cite{Saxena:2000}, the overall complexity
 of  mean-field analysis of
 the free energy~(\ref{eq1}) can be avoided by performing an
Ising-like
 description: $\vec{{\cal{M}}} = (0,0,{\cal{M}})$,
 where $ {\cal{M}} = \pm |{\cal{M}}|$ is the
 magnetization along the ``$z$-axis." Further, because of
 the equivalence of the two physical states $(\pm \cal{M})$ the thermodynamic
 analysis can be performed within the ``gauge" ${\cal{M}} \geq 0$.
 But this stage of consideration can also be achieved without
 the help of crystal anisotropy arguments. When
  the magnetic order  has a continuous symmetry one may take
advantage of the symmetry of  model~(\ref{eq1}) and avoid the
consideration of
 equivalent thermodynamic states that occur as a result of
 the respective symmetry breaking at the phase transition point but have no
 effect on thermodynamics of the system.
 In the isotropic system one may again choose a gauge,
 in which the magnetization vector
 has the same direction as  $z$-axis ($|\vec{{\cal{M}}}| =
{\cal{M}}_z
 \equiv {\cal{M}}$) and this will not influence the generality of
 thermodynamic analysis.

With the help of  convenient notations, $\varphi_j =b^{1/4}\psi_j$,
$\varphi_j = \phi_j\mbox{exp}(\theta_j)$, $M = \beta^{1/4}{\cal{M}}$,
$\gamma= \gamma_0/ (b^2\beta)^{1/4}$,
 $r = a_s/\sqrt{b}$, $t = a_f/\sqrt{\beta}$,
and neglecting the first term in~(\ref{eq1}) the free energy
 becomes
\begin{equation}
\label{eq2}
f =  r|\phi|^2 + \frac{1}{2}|\phi|^4  +
tM^2 + \frac{1}{2}M^4 +2\gamma M\phi_1\phi_2\mbox{sin}\theta\;,
\end{equation}
where $\phi^2 = (\phi_1^2 + \phi_2^2 + \phi_3^2)$, and $\theta =
 (\theta_2-\theta_1)$.

The possible (stable, metastable and unstable)
 phases are given in Table 1 together with the respective
existence and stability conditions. The stability conditions define the
 domain of the phase diagram where the respective phase is either stable
or metastable~\cite{Uzunov:1993}. The normal (disordered) phase,
denoted in Table 1 by $N$ always exists (for all temperatures $T \geq
0)$ but is stable for $t >0$, $r > 0$. The superconductivity phase
denoted in Table 1 by SC1 is unstable. The same is valid for the phase
of coexistence of ferromagnetism and superconductivity denoted in Table
1 by CO2. The N-phase, the ferromagnetic phase (FM), the
superconducting phases (SC1-3) and two of the phases of coexistence
(CO1-3) are generic phases because they appear also in the decoupled
case $(\gamma\equiv 0)$. When the $M\psi$--coupling is not present, the
phases SC1-3 are identical and represented by the order parameter
$\varphi$ where the components $\varphi_j$ participate on  equal
footing. The asterisk attached to the stability condition of ``the
second superconductivity phase",(SC2), indicates that our analysis is
insufficient to determine whether this phase corresponds to a minimum
of the free energy. As we shall see later
 the phase SC2, as
 well as the other two purely superconducting phases and the coexistence phase
 CO1, have no chance to become
 stable for $\gamma \neq 0$. This is so, because the non-generic phase of coexistence of
 superconductivity and ferromagnetism (FS in Table 1), which does
 not exist for $\gamma = 0$ is stable and
 has a lower free energy in their domain of stability.
\small

TABLE 1. Phases and their existence and stability properties
($k = 0, \pm 1,...)$.\\
\begin{tabular}{|l|l|l|l|} \hline \hline
\small Phase & order parameter & existence & stability domain \\
\hline N & $\phi_j = M = 0$ & always & $t > 0, r > 0$ \\ \hline
 FM & $\phi_j = 0$, $M^2 = -t$& $t < 0$& $r>0$, $r^2 > \gamma^2t$\\ \hline
SC1 & $\phi_1=M=0$, $\phi^2 = -r$ & $r<0$ & unstable  \\ \hline SC2 &
$\phi^2 = -r$, $\theta = \pi k$, $M = 0$ & $r<0$ & $(t > 0)^*$
 \\ \hline
SC3 & $\phi_1=\phi_2=M=0$, $\phi^2_3 = -r$ & $r<0$
&$r<0$, $t>0$\\ \hline
CO1 &$\phi_1= \phi_2=0$, $\phi^2_3 = -r$, $M^2=-t$&$r<0$, $t<0$ &
$r<0$, $t<0$ \\ \hline
CO2 &$\phi_1=0$, $\phi^2 = -r$, $\theta=\theta_2=\pi k$, $M^2=-t$&
$r<0$, $t<0$ &  unstable \\ \hline
FS & $2\phi_1^2 = 2\phi_2^2 = \phi^2 = -r + \gamma M$, $\phi_3 = 0$ &
$\gamma M > r$ &  $3M^2>(-t +\gamma^2/2)$ \\
& $\theta= 2\pi(k - 1/4) $, $\gamma r = (\gamma^2-2t)M - 2M^3$ & & $M > 0$ \\
 \hline \hline
\end{tabular}

\normalsize
We have outlined the domain in the ($t$, $r$) plane where
the FS phase exists and is a
 minimum of the free
energy. For $r < 0$ the third-degree algebraic equation  for $M$ (see
Table 1) and the existence and stability conditions are satisfied for
any $M \geq 0$ provided $t \geq \gamma^2 $. For $ t < \gamma^2$ the
condition $M \geq M_0$ have to be fulfilled, here the value
 $M_0 = (-t + \gamma^2/2)^{1/2}$ of $M$ is obtained from $r(M_0) = 0$. Thus
for $r = 0$ the N-phase is stable for
 $t \geq \gamma^2/2$, on the other hand FS is stable for $t \leq \gamma^2/2$.
For $r > 0$, the requirement for the stability of FS leads to the
inequalities
\begin{equation}
\label{eq3}
 {\mbox{max}}\left(\frac{r}{\gamma}, M_m\right) < M < M_0\;,
\end{equation}
where $M_m = (M_0/\sqrt{3})$ and $M_0$ should be the positive solution
of the third-degree equation of state from Table~1; $M_m > 0$ gives a
maximum of the function $r(M)$.

\begin{figure}
\begin{center}
\epsfig{file=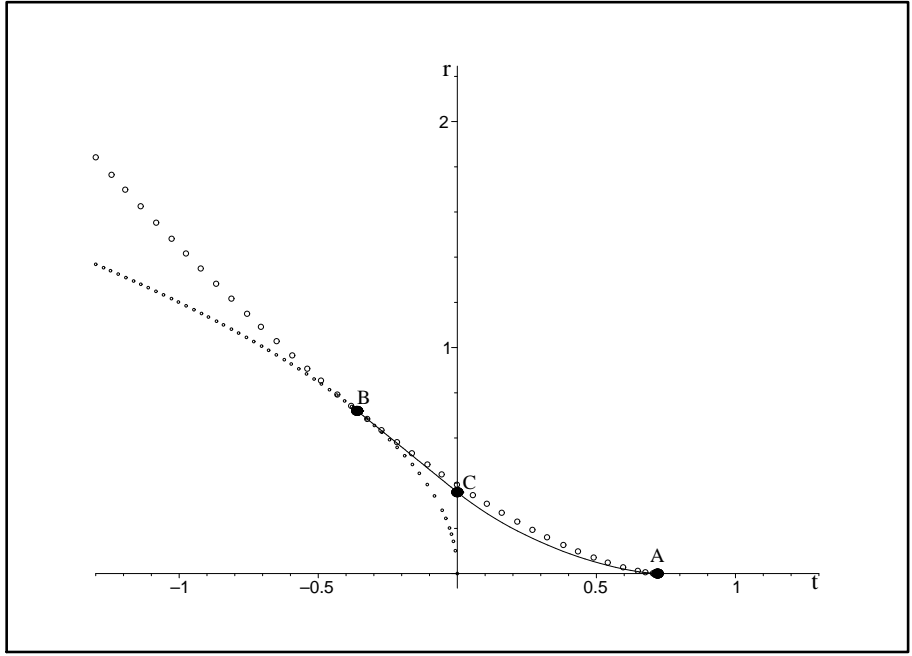,angle=-90, width=12cm}\\
\end{center}
\caption{The phase diagram in the plane ($t$, $r$) with two tricritical
points (A and B) and a triple point $C$; $\gamma = 1.2$.}
\label{NEW11.fig}
\end{figure}

The further analysis leads to the existence and  stability domain of FS
below the line AB given by circles (see Fig.~1). In Fig.~1 the curve of
circles starts from the point A with coordinates ($\gamma^2/2$, $0$)
and touches two other (solid and dotted) curves at the point B with
coordinates ($-\gamma^2/4$, $\gamma^2/2$).  Line of
 circles represents the function
$r(M_m) \equiv r_m(t)$ where
\begin{equation}
\label{eq4}
 r_m(t) = \frac{4}{3\sqrt{3}\gamma} \left (\frac{\gamma^2}{2} -
 t\right)^{3/2}.
\end{equation}
Dotted line is given by $r_e(t) = \gamma\sqrt{|t|}$. The inequality $r
< r_m(t)$ is a condition for the stability of FS, whereas the
inequality $r \leq r_e(t)$ for $ (-t) \geq \gamma^2/4$ is a condition
for the existence of FS as a solution of the respective equation of
state. This existence condition for FS has been obtained from $\gamma M
> r$ (see Table 1).

In the region on the left of the point B in Fig.~1, the FS phase
satisfies the existence condition $\gamma M > r$ only
 below the dotted line. In the domain confined between the lines of circles
 and the dotted
line on the left of the point B the stability condition for FS is
satisfied but the existence condition is broken. The inequality $r \geq
r_e(t)$ is the stability condition of FM for $ 0 \leq (-t) \leq
\gamma^2/4$. For $(-t) > \gamma^2/4$ the FM phase is stable for all $r
\geq r_e(t)$.

In the region confined by the line of circles AB, the dotted line for $
0 < (-t) < \gamma^2/4$, and the $t-$axis, the phases N, FS and FM have
an overlap of stability domains. The same problem exists for FS and the
SC phases in the second quadrant and for the phases FS and CO1 in the
third quadrant of the plane ($t$, $r$). The comparison of the
respective free energies for $r < 0$ shows that the stable phase is FS
whereas the other phases are metastable within their domains of
stability.

The part of
 the $t$-axis given by $r=0$ and $t > \gamma^2/2$
 is a phase transition line of second order
which describes the N-FS transition. The same transition
 for $0 < t < \gamma^2/2$ is represented by the solid line AC which
is the equilibrium transition line of a first order phase transition.
This equilibrium transition curve is given by the function
\begin{equation}
\label{eq5}
 r_{eq}(t) =
\frac{1}{4}\left[3\gamma - \left(\gamma^2 + 16t
\right))^{1/2}\right]M_{eq}(t),
\end{equation}
where
  \begin{equation}
\label{eq6}
 M_{eq}(t) =
\frac{1}{2\sqrt{2}}\left[\gamma^2 - 8t + \gamma\left(\gamma^2 +
 16t \right)^{1/2}\right]^{1/2}
\end{equation}
is the equilibrium value (jump) of the magnetization. The order of
the
N-FS
 transition changes at the tricritical point A.

The domain above the solid line AC and below the line of circles for $
t > 0$ is the region of a possible
 overheating of FS.
The domain of overcooling of the N-phase is confined by the solid line AC and
the axes ($t > 0$, $r >0$). At the triple point C with coordinates
 [0, $r_{eq}(0) = \gamma^2/4$]
the phases N, FM, and FS
coexist. For $t < 0$ the straight line
\begin{equation}
\label{eq7}
r_{eq}^* (t) =  \frac{\gamma^2}{4} + |t|,\;\;\;\;\;\; -\gamma^2/4 < t < 0,
\end{equation}
describes the extension of the equilibrium phase transition line of the
N-FS first order transition to negative values of $t$.
 For $t < (-\gamma^2/4)$
 the equilibrium phase transition FM-FS is of second order and is
given by the dotted line on the left of the point B (the second
tricritical point in this phase diagram). Along the first order
transition line
 $r_{eq}^{\ast}(t)$ given by~(\ref{eq8}) the equilibrium value
 of $M$ is $M_{eq} =\gamma/2$,  which
implies an equilibrium order parameter jump at the FM-FS transition equal to
($\gamma/2 - \sqrt{|t|}$). On the dotted line of the second order
FM-FS
 transition the equilibrium value
of $M$ is equal to that of the FM phase ($M_{eq} = \sqrt{|t|}$). Note,
that the FS phase does not exists below $T_s$ and this seems to be a
disadvantage of the model~(\ref{eq1}).

The equilibrium phase transition lines of the FM-FS and N-FS transition
lines in Fig.~1 can be expressed by the respective equilibrium phase
transition temperatures $T_{eq}$ defined by the equations $r_e =
r(T_{eq})$, $r_{eq} = r(T_{eq})$, $r^{\ast}_{eq} = r(T_{eq})$, and with
the help of the relation $M_{eq} = M(T_{eq})$. This leads to some
limitations on the possible variations of the parameters of the theory.
 For example, the critical temperature
($T_{eq} \equiv T_c$) of the FM-FS transition of second order
 ($\gamma^2/4 < -t$)  is obtained in the form
$T_{c} = (T_s + 4\pi J{\cal{M}}/\alpha_s)$,
or, using ${\cal{M}} = (-a_f/\beta)^{1/2}$,
\begin{equation}
\label{eq8}
T_{c} = T_s -\frac{T^{\ast}}{2} +
\left[ \left(\frac{T^{\ast}}{2}\right)^2 + T^{\ast}(T_f-T_s)\right]^{1/2}\;,
\end{equation}
where $T_f > T_s$, and $T^{\ast} = (4\pi J)^2\alpha_f/\alpha_s^2\beta$ is
 a characteristic temperature of the model~(\ref{eq1}). The
 investigation of the conditions for the validity of~(\ref{eq8}) leads to the
 conclusion that the
FM-FS continuous phase transition (at $\gamma^2 < -t)$ is possible only
if the following condition is satisfied:
\begin{equation}
\label{eq9}
T_{f} - T_s > \ = (\varsigma + \sqrt{\varsigma})T^{\ast}\;,
\end{equation}
where $\varsigma = \beta\alpha_s^2/4b\alpha_f^2$.
 This means that the second
order FM-FS transition should disappear for sufficiently large
$M\psi$--coupling. Such a condition does not exist for the first order
transitions FM-FS and N-FS.

Taking into  account the first term in the free energy~(\ref{eq1})
should lead to a depression of the equilibrium transition temperature.
As the magnetization increases with the decrease of the temperature,
the vortex state should occur at temperatures which are lower than
 the equilibrium temperature $T_{eq}$ of
the homogeneous (Meissner) state. For example,
the critical temperature ($\tilde{T}_c$)
 corresponding to the inhomogeneous (vortex) phase
of FS-type has been evaluated~\cite{Walker:2002} to be
 lower than the critical temperature ~(\ref{eq8}): $(T_c - \tilde{T}_c) =
4\pi \mu_B{\cal{M}}/\alpha_s$ ($\mu_B = |e|\hbar/2mc$ - Bohr magneton).
 For $J \gg \mu_B$, we have
 $T_c \approx \tilde{T}_c$. Finally, let us emphasize that a more reliable
description of these phenomena, in particular, of the thermodynamic
behaviour of the FS phase at relatively large values of ${\cal{M}}$ can
be performed if an additional term of type ${\cal{M}}^2|\psi|^2$ is
included in the model~(\ref{eq1}).

{\bf Acknowledgments:} DIU thanks the hospitality of MPI-PKS-Dresden.
Financial support through {\em Scenet} (Parma) and collaborative
project with JINR-Dubna is also acknowledged.


\newpage
\begin{thebibliography}{ll}
\bibitem{Saxena:2000}
S. S. Saxena, P. Agarwal, K. Ahilan, F. M. Grosche, R. K. W. Haselwimmer,
M.J. Steiner, E. Pugh, I. R. Walker, S.R. Julian, P. Monthoux, G. G. Lonzarich,
A. Huxley. I. Sheikin, D. Braithwaite, and J. Flouquet,  Nature
(London)
 406 (2000) 587.
\bibitem{Huxley:2001}
A. Huxley, I. Sheikin, E. Ressouche, N. Kernavanois, D. Braithwaite,
R. Calemczuk, and J. Flouquet, Phys. Rev. B 63 (2001) 144519-1.
\bibitem{Tateiwa:2001}
N. Tateiwa, T. C. Kobayashi, K. Hanazono, A. Amaya, Y. Haga. R. Settai, and Y.
Onuki, J. Phys. Condensed Matter, 13 (2001) L17.
\bibitem{Pfleiderer:2001}
C. Pfleiderer, M. Uhlatz, S. M. Hayden, R. Vollmer, H. v.
L\"ohneysen,
N. R. Berhoeft, and G. G. Lonzarich, Nature (London) 412 (2001) 58.
\bibitem{Aoki:2001}
D. Aoki, A. Huxley, E. Ressouche, D. Braithwaite, J. Flouquet, J-P.. Brison,
E. Lhotel, and C. Paulsen, Nature (London) 413 (2001) 613.
\bibitem{Shimizu:2001}
K. Shimizu, T. Kimura, S. Furomoto, K. Takeda, K. Kontani, Y. Onuki
 and K. Amaya, Nature (London) 412 (2001) 316.
\bibitem{Volovik:1985}
G. E. Volovik and L. P. Gor'kov, Sov. Phys. JETP
 61 (1985) 843 [Zh. Eksp. Teor. Fiz. 88 (1985) 1412].
\bibitem{Sigrist:1991}
M. Sigrist and K. Ueda, Rev. Mod. Phys.
  63 (1991) 239.
\bibitem{Mineev:1999}
V. P. Mineev, K. V. Samokhin, Introduction to Unconventional
Superconductivity (Gordon and Breach, Amsterdam, 1999).
\bibitem{Saxena:2001}
S. S. Saxena and P. B. Littlewood, Nature (London) 412 (2001) 290.
\bibitem{Machida:2001}
K. Machida and T. Ohmi, Phys. Rev. Lett. 86 (2001) 850.
\bibitem{Walker:2002}
M. B. Walker and K. V. Samokhin, {\em Phys. Rev. Lett.}
  88 (2002) 204001-1.
\bibitem{Uzunov:1993}
D. I. Uzunov, Theory of Critical Phenomena (World Scientific, Singapore, 1993).
\bibitem{Blagoeva:1990}
E. J. Blagoeva, G. Busiello, L. De Cesare, Y. T. Millev, I. Rabuffo, and D. I.
Uzunov, Phys. Rev. B 40 (1990) 7321.
\bibitem{Uzunov:1990}
D. I. Uzunov, in: Advances in Theoretical Physics, ed. by E.
Caianiello
(World Scientific, Singapore, 1990) p. 96.
\bibitem{Gufan:1980}
Yu. M. Gufan and V. I. Torgashev, Sov. Phys. Solid State 22 (1980)
951
[Fiz. Tv. Tela (Leningrad) 22 (1980) 1629].
\bibitem{Gufan:1986}
Yu. M. Gufan, E. I. Kut'in, V. L. Lorman, and E. N. Sidorenko,
  Sov. Phys. Solid State 29 (1987) 432
[Fiz. Tv. Tela (Leningrad) 29 (1986) 756].
\end{thebibliography}
 \end{document}